\documentclass[10pt, conference, compsocconf]{IEEEtran}

\usepackage{epsfig}
\usepackage{epstopdf}
\usepackage{pdfpages}
\usepackage{graphicx}
\usepackage{url}
\usepackage{changebar}
\usepackage{xcolor}
\usepackage{soul}
\usepackage[caption = false]{subfig}
\usepackage{url}
\usepackage{xspace}
\usepackage{amssymb}

\usepackage{listings}
\usepackage{fancyvrb}

\definecolor{dkgreen}{rgb}{0,0.6,0}
\definecolor{gray}{rgb}{0.5,0.5,0.5}
\definecolor{mauve}{rgb}{0.58,0,0.82}

\ifCLASSINFOpdf
\else
\fi
\begin{document}
%\title{A Portable Bandwidth-saving MPI Broadcast Approach}
\title{A Bandwidth-saving Optimization for MPI Broadcast Collective Operation}


\author{\IEEEauthorblockN{Huan Zhou}
\IEEEauthorblockA{High Performance Computing Center Stuttgart (HLRS)\\
University of Stuttgart, Germany}
\and 
\IEEEauthorblockN{Vladimir Marjanovic}
\IEEEauthorblockA{High Performance Computing Center Stuttgart (HLRS)\\
University of Stuttgart, Germany}
\and
\IEEEauthorblockN{Christoph Niethammer}
\IEEEauthorblockA{High Performance Computing Center Stuttgart (HLRS)\\
University of Stuttgart, Germany}
\and
\IEEEauthorblockN{Jos\'e Gracia}
\IEEEauthorblockA{High Performance Computing Center Stuttgart (HLRS)\\
University of Stuttgart, Germany}

%\and
%\IEEEauthorblockN{Authors Name/s per 2nd Affiliation (Author)}
%\IEEEauthorblockA{line 1 (of Affiliation): dept. name of organization\\
%line 2: name of organization, acronyms acceptable\\
%line 3: City, Country\\
%line 4: Email: name@xyz.com}
}


%Telephone: (800) 555--1212, Fax: (888) 555--1212}


\maketitle

\begin{abstract}
The efficiency and scalability of MPI collective operations, in particular
the broadcast operation, plays an integral part in high performance
computing applications.
%Most of the research works in the MPI collective communication operations, 
%particularly in MPI broadcast, are carried out to tune it
%for certain platform according to either the topology or some attracting hardware-supported features. 
MPICH, as one of the contemporary widely-used MPI software stacks,
implements the broadcast operation based on point-to-point operation.
Depending on the parameters, such as
message size and process count, the library chooses to use
different algorithms, as for instance binomial dissemination, recursive-doubling exchange or ring all-to-all broadcast (allgather).
However, the existing broadcast design in latest release of MPICH does
not provide good performance for large messages (\textit{lmsg}) or 
medium messages with non-power-of-two process counts (\textit{mmsg-npof2}) 
due to the inner suboptimal ring allgather algorithm. In this paper, based on the native broadcast design in MPICH, 
we propose a tuned broadcast approach with bandwidth-saving in mind catering to 
the case of \textit{lmsg} and \textit{mmsg-npof2}. 
Several comparisons of the native and tuned broadcast designs are made
for different data sizes and program sizes on Cray XC40 cluster.
The results show that the performance of the tuned broadcast design can get improved by a range from 2\% to 54\% for \textit{lmsg}
and \textit{mmsg-npof2} in terms of user-level testing.

\end{abstract}

\begin{IEEEkeywords}
MPICH, Broadcast, Bandwidth-saving

\end{IEEEkeywords}

% \fi
\IEEEpeerreviewmaketitle

\section{Introduction}
\label{introduce}

The message passing interface (MPI)~\cite{mpi} standard provides a flexible and portable environment for developing high performance 
parallel applications on different platforms. Since the release of the first version of MPI, it has become a 
very flexible communication layer providing mechanisms for both point-to-point and collective operations.

The MPI standard specifies various types of collective operations~\cite{mpi3} such as 
\textit{All-to-All} ({\em MPI$\_$Allgather}, {\em MPI$\_$Allscatter} and {\em MPI$\_$Alltoall}), 
\textit{All-to-One} ({\em MPI$\_$Gather} and {\em MPI$\_$Reduce}) and \textit{One-to-All} 
(dissemination: {\em MPI$\_$Bcast} and {\em MPI$\_$Scatter}). 
Many scientific applications use collective communications to synchronize or 
exchange data. 

Noticeably, collective communication is a critical and also frequently used component in MPI.
In particular, MPI broadcast, as the typical \textit{One-to-All} interface,
is an essential interface widely used in many scientific fields, for instance,
matrix multiplication, basic linear algebra~\cite{hpl} and so on.
Besides, a profiling study~\cite{LS-DYNA} shows that
the efficiency of MPI broadcast operations can have a remarkable impact on the overall 
LS-DYNA~\cite{LS-DYNA} performance. Therefore, 
the MPI implementors are willing to put great efforts on the optimization of MPI broadcast implementation.

MPICH~\cite{mpich-overview} is a high-performance, freely-available and portable implementation of the MPI
and MPICH3 supports the latest version of the MPI standard -- MPI-3.
In addition, A vast majority of other MPI implementations, including IBM MPI, Intel MPI, Cray MPI, OSU MVAPICH/MVAPICH2 and so forth, are
derived from MPICH.       
According to the new Top500 Supercomputer report~\cite{top500}, of the top 10 Supercomputers as of November 2014,
90\% are exclusively using MPICH or its derivatives.

Multi-core processor~\cite{multicore} emerges to speed up 
the computation capability of processor through
performing the workload among multiple cores concurrently.
With the advantage of \mbox{multi-core} processor, it has been commonly deployed in the nowadays computation
clusters.
Accordingly, we can basically exploit two communication 
levels -- intra-node and inter-node~\cite{bcast-tu} on multi-core clusters to analyze the broadcast algorithm in MPICH3.

MPICH has already highly tuned the implementation of broadcast operation in a way of
using multiple algorithms based on the message size and process count~\cite{mpich-opt} and 
such implementation design is still employed in MPICH3~\cite{mpich-sc}.
However, there is still considerable room left at optimizing broadcast especially
when transferring the long messages (\textit{lmsg}), or medium messages but with non-power-of-two process counts (\textit{mmsg-npof2}).
The occurrence of non-power-of-two processes can be due to explicit
user request at  job-launching, particularly on systems where the
core count per node is already non-power-of-two,  or due to splitting on 
the communicator in the applications.

For \textit{lmsg}, MPICH3 adopts a \textit{scatter-ring-allgather} approach, where
the source data to be broadcast is first divided up and scattered among the processes
following a binomial tree from root. The scattered data are then collected back
to all processes as a ring allgather operation.
For \textit{mmsg-npof2}, the implementation is of multi-core awareness, where we first perform an intra-node broadcast on the node of root following a binomial tree,
then perform the inter-node broadcast
by using the \textit{scatter-ring-allgather}, finally perform the intra-node broadcast on all nodes except for the node of root again
following a binomial tree. 
However, the suboptimal ring allgather operation potentially involves verbose data transmissions which actually can be avoided.
Further, for \textit{lmsg}, it is to be noted that we need put considerable emphasis on the usage of bandwidth, which 
implies that the number of useless data transmissions should be minimized aiming to save bandwidth use.
This is an interesting subject that is however
easily overlooked by most active MPI researchers.

Therefore, in this paper, we investigated in-depth the methodology of designing an optimized broadcast algorithm
(\textit{scatter-ring-allgather})
particularly for \textit{lmsg} and \textit{mmsg-npof2} by tuning the suboptimal ring allgather design.
%based on the original broadcast design in MPICH3 from the algorithm perspective.
The scalability and portability of the optimized broadcast algorithm
can be maintained since such optimization is not bound to particular platform or
special features in hardware. 
%In order to compare the native
%and tuned broadcast design, for simplicity,

We have implemented the native and tuned \textit{scatter-ring-allgather} algorithms without multi-core awareness
on the user level. Hence we can compare the native and tuned \textit{scatter-ring-allgather} algorithms for \textit{lmsg} and 
\textit{mmsg-npof2} case from a broader perspective, which allows us to observe
their performance difference when the inter-node or intra-node data transmissions are involved. 
Here on, all references to the user-level implementation of the native and tuned \textit{scatter-ring-allgather} algorithms free of multi-core awareness
refer to {\em MPI\_Bcast\_native} and {\em MPI\_Bcast\_opt}, respectively.
%In addition, the theoretical analysis based on the simple MPI performance model -- LogP~\cite{logp} is presented 
%to show the potential benefit of our design. 
We conduct a series of experiments on Cray XC40 to indicate
that the tuned design can improve the bandwidth performance of
broadcast operation by a range from 2\% to 54\% for 
\textit{lmsg} and \textit{mmsg-npof2} in terms of the user-level testing.

The rest of this paper is organized as follows: In Section \ref{relatedwork} we discuss related work.
In Section \ref{background}, we provide an overview of the native \textit{scatter-ring-allgather}
algorithm ({\em MPI\_Bcast\_native}). In Section \ref{design}, 
the tuned design of \textit{scatter-ring-allgather} algorithm ({\em MPI\_Bcast\_opt}) is described and 
explained with the pseudo-code as well. We evaluate the tuned design and compare it with {\em MPI$\_$Bcast\_native} in Section \ref{evaluation}.
Conclusion is presented in Section \ref{conclusion}.

\section{Related Work}
\label{relatedwork}
There have been many careful studies about the optimization of broadcast implementation
targeting for the specific interconnects. 
Two papers~\cite{bcast-multicast-liu,bcast-hoefler}
focus on the InfiniBand clusters with hardware-supported multicast,
which can improve the overall performance 
of broadcast significantly and however are closely dependent of the underlying interconnects.
Additionally,
A study~\cite{conf/europar/bcast_sdn} demonstrates that
the broadcast performance can get improved on the Software-Designed network,
which is of controllability.
The MPI broadcast operations get optimized 
as a result of the network hardware acceleration for 
broadcast provided by the Blue Gene/Q, shown in paper~\cite{journals/ijhpca/bcast_ibm}.
However, those optimized designs will show poor portability when applying them on other networks without multicast support,
controllability or dedicated hardware acceleration.

\section{Background}
\label{background}
\subsection{Overview of MPI\_Bcast\_native}
\label{binomial}
MPI standard specifies that the broadcast operation should disseminate a message from a root process to other processes 
in a communication group. MPI broadcast is a blocking operation, which means all processes are ready
to use the received data after the broadcast operation is successfully returned.

The algorithm that is generally used by {\em MPI\_Bcast\_native} for \textit{lmsg} and \textit{mmsg-npof2}
is the combination of a binomial scatter and a suboptimal ring allgather operation.
We assume that there are $P$ processes participating in the broadcast operation.
In theory, {\em MPI\_Bcast\_native} first uses a binomial scatter to make process $i$ get the $i$-th block of data from root
(Section \ref{introduce} mentioned that the data source should be divided up into
$P$ pieces of data block before the scatter operation), then invokes a ($P-1$)-step enclosed-ring allgather operation.

\begin{figure}
 \begin{center}
 \includegraphics[scale=0.3]{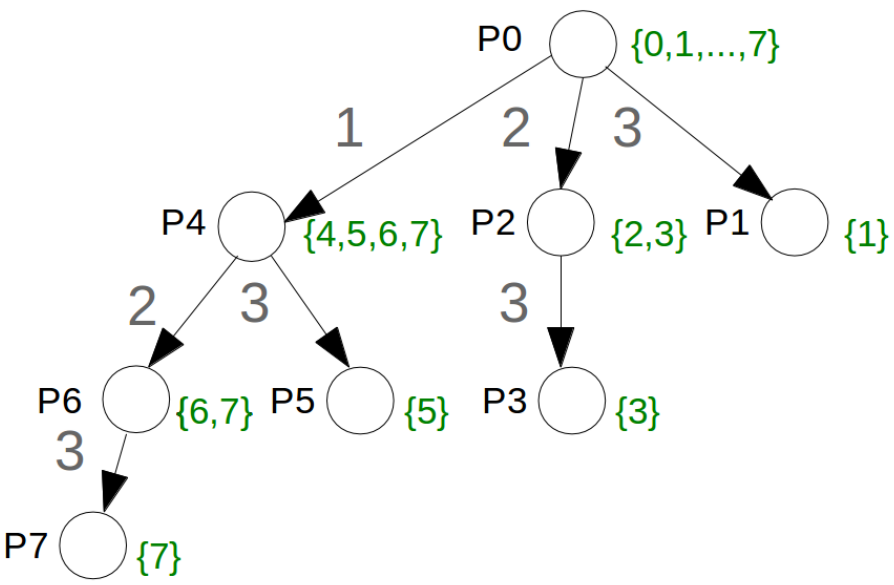}
\end{center}
\caption{Schematic for the binomial scatter operation in the case of 8 processes}
\label{scatter}
\end{figure}

\begin{figure}
 \begin{center}
 \includegraphics[scale=0.22]{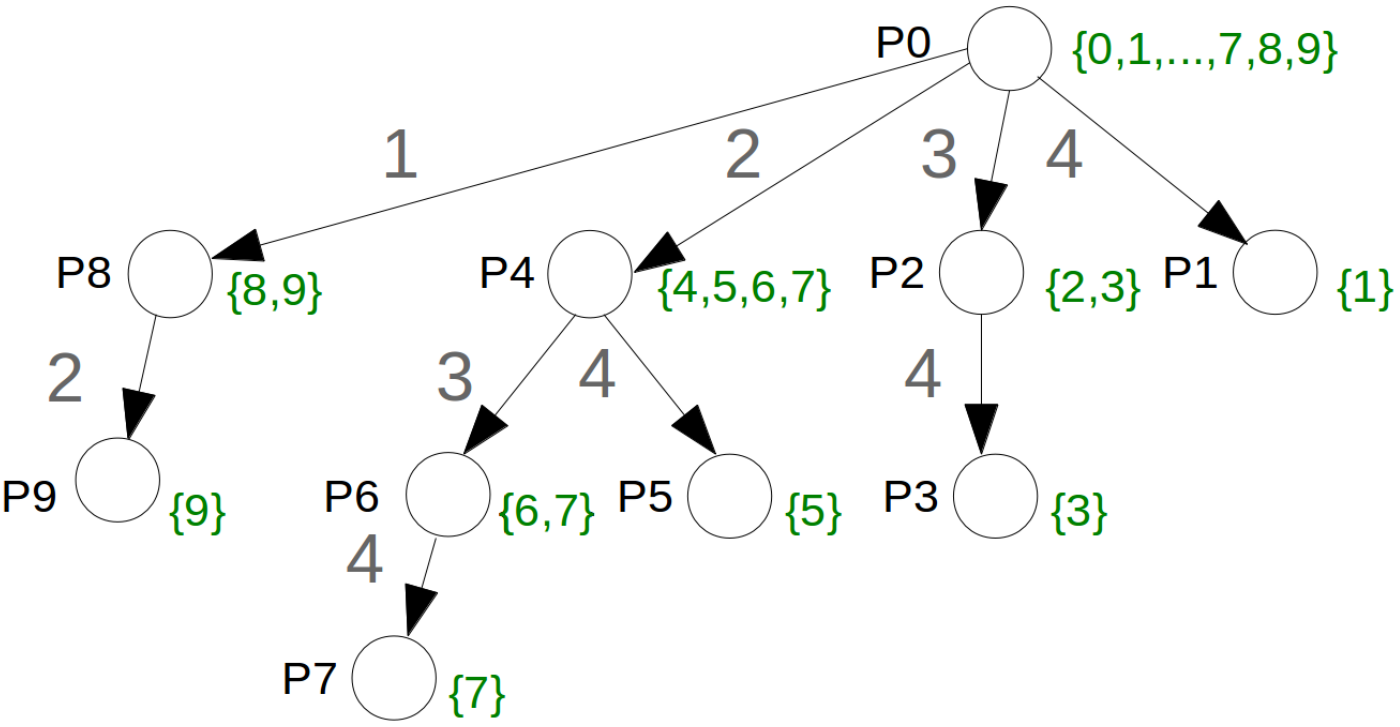}
\end{center}
\caption{Schematic for the binomial scatter operation in the case of 10 processes}
\label{npof2-scatter}
\end{figure}

On the one hand, we explain how the scatter operation
proceeds following a binomial tree from the root $0$ for a power-of-two number of processes -- 8 processes, shown in Figure
\ref{scatter}.
The root $0$ divides the source data into $8$ chunks,
where each chunk will be marked as a non-negative number $i$ and supposed to be transmitted to the corresponding process $pi$.
In this way the source data in root $0$ consists of 8 chunks of data marked with $0$, $1$, ..., $7$ sequentially.
In the first step, root 0 sends the chunk set of $\{$4,5,6,7$\}$ to process 4, then 
a sub-tree is spawned, as process $p4$ be the root. 
In the following step, we spawn two new sub-trees, as process 2 and 6 be the root respectively.
Finally each process is able to get the corresponding data after the third step
and a binomial tree spanning 8 processes is completed. 
Generally all processes
can get the corresponding data in $log_{2}P$ steps for power-of-two processes.
On the other hand, we, take 10 processes for example, describe 
the generation of a binomial scatter tree for non-power-of-two processes.
The scatter path keeps the same as the path in Figure \ref{scatter} except that 
an additional branch, as process $8$ be the root, is spawned.
In this case the entire scatter operation finishes in $\lceil log_{2}P \rceil$ steps,
shown in Figure \ref{npof2-scatter}.
According to the above two figures, 
we can conclude that practically not only does each non-leaf node $pi$ in the binomial scatter tree own its corresponding chunk of data marked with $i$, 
but it also provides all data chunks for its descendant.

\begin{figure}
\begin{center}
 \includegraphics[width=0.48\textwidth,height=0.28\textheight]{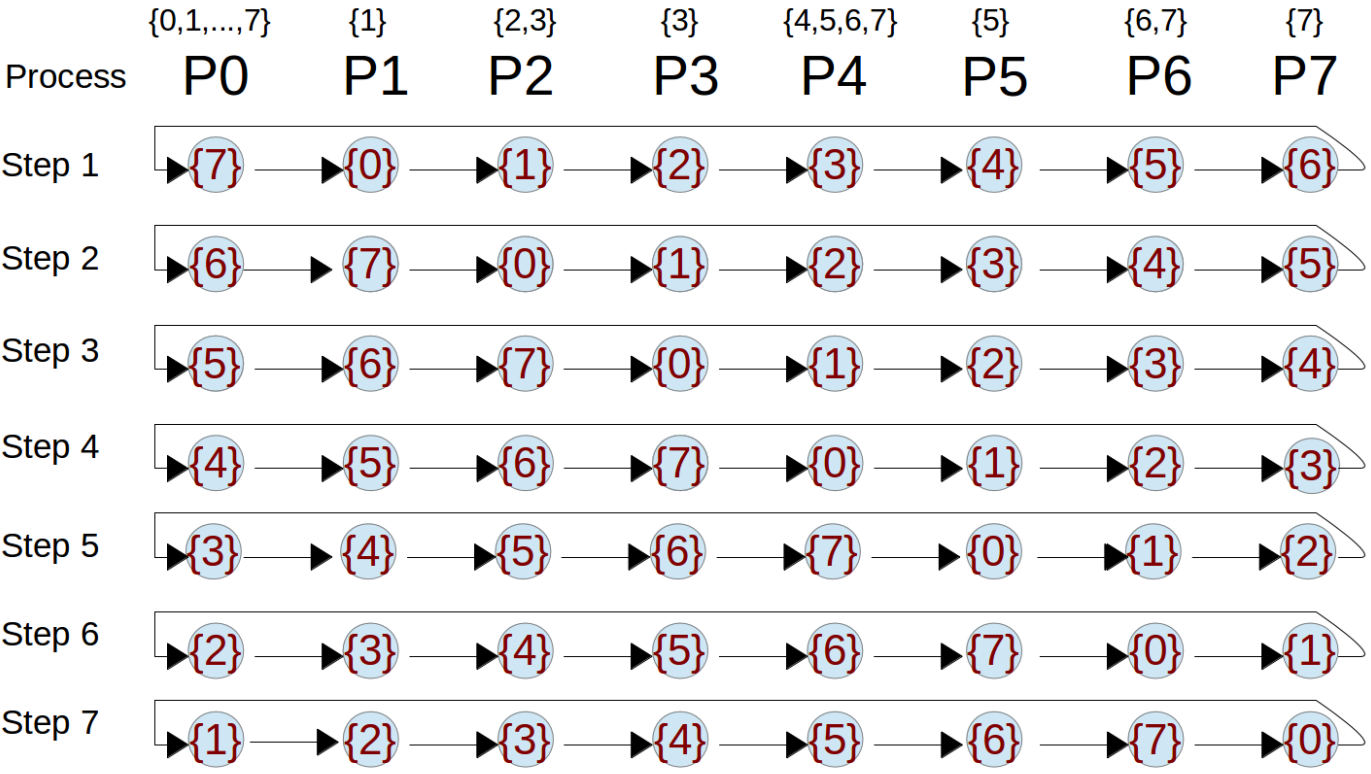}
\end{center}
\caption{Enclosed ring for the original allgather algorithm in the case of 8 processes}
\label{enclosed-ring}
\end{figure}

Figure \ref{enclosed-ring} shows us the suboptimal ring allgather operation for 8 processes,
process $i$ sends its contribution to process $i+1$ and meantime receives the contribution from process $i-1$ (with wrap-around)
in the first step and from the second to seventh step each process $i$ forwards to process $i+1$ the data it received from process $i-1$ during the previous 
step. The set that is listed above each process indicates all
data chunks that it owns in reality after the binomial scatter operation, illustrated in Figure \ref{scatter}.
However we can find that this allgather algorithm is carried out in an enclosed ring manner, 
where each process $pi$ pretends to only own the $i$-th 
data chunk and thus repeatedly receives the data chunks that already existed in it. This obviously leads to a large amount of useless data transmissions in each step.
For the non-power-of-two processes, the algorithm of ring allgather operation goes the same way as that for power-of-two processes.
Therefore, generally there are totally data transmissions of $P\times(P-1)$ in this suboptimal ring allgather operation with 
data transmissions of $P$ at each step.

\section{Designing {\em MPI\_Bcast\_opt}}
\label{design}
In this section, we describe the new broadcast algorithm for 
\textit{lmsg} and \textit{mmsg-npof2} 
and its potential advantage over the original algorithm used by MPICH3,
which has been elaborated in Section \ref{background}.

\begin{figure}
\begin{center}
 \includegraphics[width=0.48\textwidth,height=0.28\textheight]{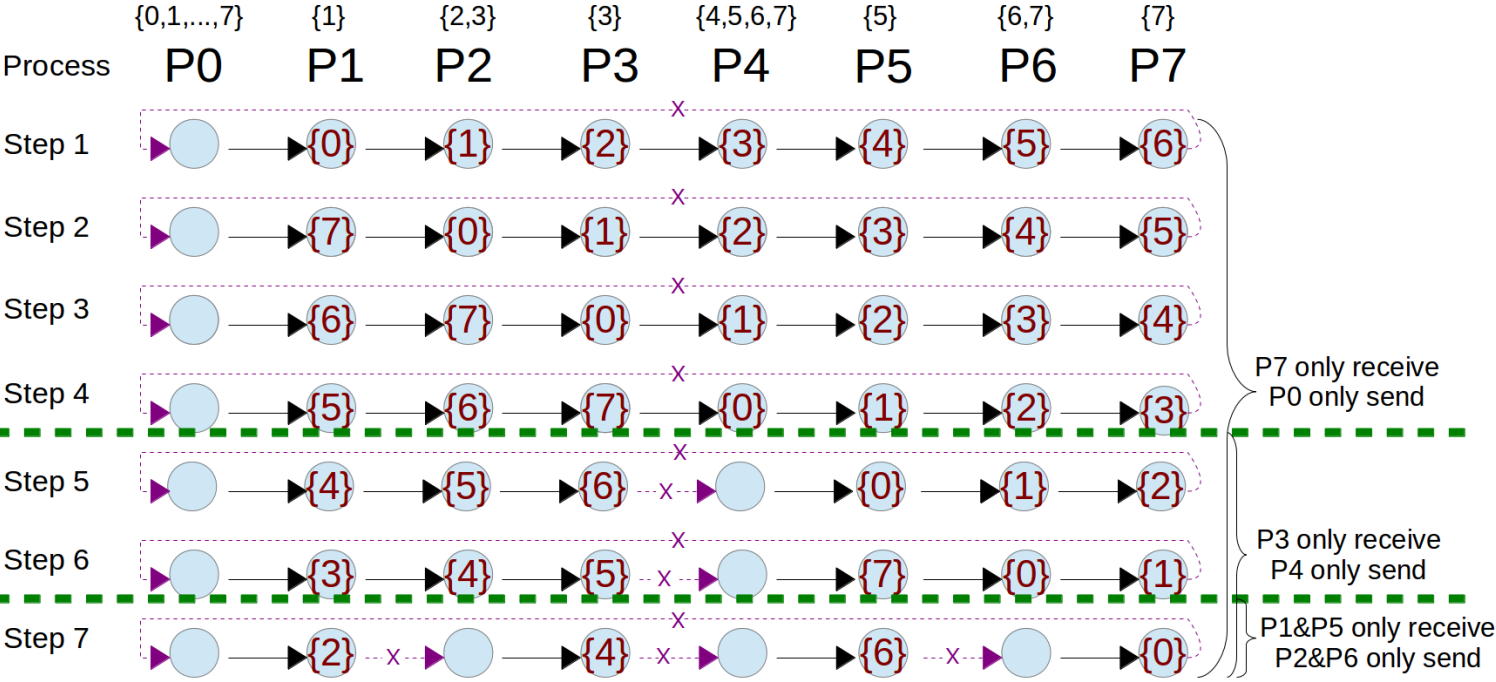}
\end{center}
\caption{Non-enclosed ring for the tuned allgather algorithm in the case of 8 processes}
\label{ring}
\end{figure}

We optimize the native broadcast algorithm in MPICH3 by tuning the suboptimal ring allgather operation in a way of letting
each process be aware of the actual data chunks that it already owned and skipping the verbose 
data transfers happening in the original ring allgather operation (see Figure \ref{enclosed-ring}).
Figure \ref{ring} shows the tuned ring allgather algorithm by
illustrating the send and receive events that happen in each step with run size of 8.

Likewise, the set shown in the top row lists all data chunks that a process has already 
owned after doing the binomial scatter operation.
Noticeably, we can see that the new ring allgather algorithm proceeds in a non-enclosed ring manner.
Process 0 is the root owning the source data and thus it does not need receive any chunk of data from process 7.
In the first four steps, process 4 gets the data chunks marked with 3,2,1 and 0 from process 3 in sequence,
which complete the receive buffer of process 4 together with its
existing chunks of data marked with 4,5,6 and 7. Therefore, from the fifth step on process 4 stop  
receiving any further data chunk from process 3. by analogy, we can see that 
process 2 and 6 collect data chunks from process 1 and 5 till the seventh step, where process 2 and 6 already gets all
the data chunks that they are lack of.
Therefore, each process only receives those missing data chunks, ignores the repeated data chunks and also terminates
in $8-1=7$ steps.
The number of message transfers in the original ring allgather algorithm is $8\times(8-1)=56$ for 8 processes,
by contrast, the tuned design can reduces it by $12$.

\begin{figure}
\begin{center}
 \includegraphics[width=0.48\textwidth,height=0.28\textheight]{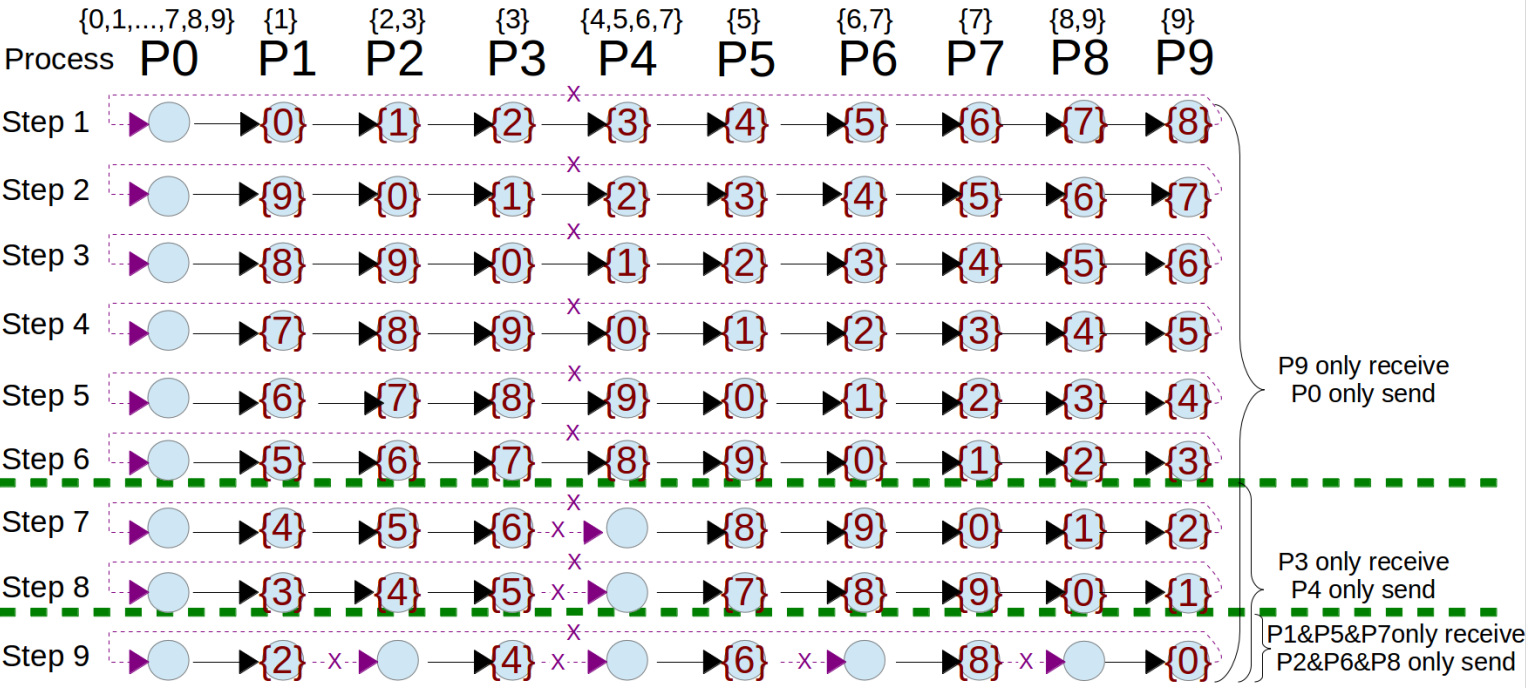}
\end{center}
\caption{Non-enclosed ring for the tuned allgather algorithm in the case of 10 processes}
\label{npof-2-ring}
\end{figure}

Figure \ref{npof-2-ring} exhibits a more complex scenario, where the tuned ring allgather algorithm
performs with 10 processes as an example of non-power-of-two. 
Process 4 stops receiving data chunks from process 3 after reaching the sixth step since it already gets 
all missing data chunks marked with 3,2,1,0,9 and 8 sequentially. Additionally, 
not only process 2 and 6 but also process 8 get the full source data chunks which are broadcast initially by root after eighth step.
In this case, the number of message transfers is $75$ and reduced by $15$
compared with $10\times(10-1)=90$ brought by the original ring allgather algorithm.

%the tuned ring allgather algorithm produces $lgP \times P/2$ less transmitted data than the native ring allgather for the power-of-two processes -- $P$,
According to Figures \ref{ring} and \ref{npof-2-ring}, we can deduce that the decrement in the amount of the transferred data will increase as the growing of the process count $P$.
Furthermore, by combining Figure \ref{enclosed-ring} and Figure \ref{ring}, we can conclude that the tuned ring allgather algorithm reduces the 
data transmission traffic to an extent using the same steps as the native ring allgather algorithm.

Intuitively, the tuned ring allgather algorithm can speedup the broadcast performance by reducing unnecessary data transmissions.
%and then alleviating total start-up overhead.
Technically, in the case of intra-node, the point-to-point operation is implemented via memory copying, which is considered to
involve the cpu-interference and buffer memory allocation, which can be minimized in the tuned ring allgather algorithm.
In the case of inter-node, the source data should be sent into network by the sender.
In addition, there is currently no high performance cluster capable to reach a ideal capability, where 
the network environment is not negatively influenced by the quantity of data transmission and has sufficient memory resource or unlimited network bandwidth.
Thus, with the tuned ring allgather algorithm not only can we save the buffer memory, but we also save bandwidth by decreasing the amount of 
messages injecting into it and therewith bring down the opportunity of network congestion. 
Accordingly, we can infer that the tuned ring allgather algorithm can bring potential performance benefit for broadcast operation
on both the two communication levels.

Pseudo-code for the tuned ring broadcast algorithm for \textit{lmsg} and \textit{mmsg-npof2} ({\em MPI$\_$Bcast$\_$opt})
is presented in listing \ref{opt-code}, where the added code is the auxiliary part helping to realize the tuned ring allgather algorithm.
First, we assume the communicator size is $P$, root divides the source data into pieces of $P$ and then scatters those pieces to the other processes following a binomial tree.
Second, each process computes out the related \textit{step}, which indicates that each process starts to either send or receive data chunks from ($P-\textit{step}+1$)-th step
on and the direction of data transmission is determined by the value of flag.
At last, unlike the native ring 
allgather algorithm,
%where all processes send and meantime receive data chunk during each step ($P$-1 steps in total) no matter what data chunks it already owns, 
the tuned ring allgather design makes all processes start the collection of
data chunks according to \textit{step} and flag in the purpose of omitting the useless data transmissions.
\begin{lstlisting}[mathescape, escapechar=', label=opt-code]
 void MPI_Bcast_opt (char *buffer,int count,int length,int root,MPI_Comm comm)
 {
 /* Get the process rank and communicator size */
 MPI_Comm_rank(comm,&rank);
 MPI_Comm_size(comm,&comm_size);
 
 /* If the process 0 is not the root, then each process needs to get the relative_rank with respect to the root */
 relative_rank=(rank>=root)?\
         rank-root:(rank-root+comm_size);
         
 /* Root devides the source data into pieces of comm_size and disseminates them to the other processes in a binomial tree */
 scatter_size=(nbytes+comm_size-1)/comm_size;
 
 /* See Figure '\ref{scatter}'&'\ref{npof2-scatter}' for details */ 
 binomial_tree(buffer,count,length,root,comm); 
 
 /* --- The tuned ring allgather algorithm --- */
 
 /* Each process computes the absolute left node and right node in the virtual ring */
 left=(comm_size+rank-1)%comm_size;
 right=(rank+1)%comm_size;
 j=rank;
 jnext=left;
 
 /* Added code: Each process calculates the '\textit{step}' based on which it decides to either send or receive data inside the ring allgarther operation */
 mask=$2^{ceil(log_{2}comm\_size)}$;
 while(mask>1){
  right_relative_rank=(relative_rank+1<comm_size)?\
          relative_rank+1:relative_rank+1-comm_size;
  if(!(right_relative_rank%mask)){
    '\textit{step}'=mask;
    if(right_relative_rank+mask>comm_size){
      '\textit{step}'=comm_size-right_relative_rank;}
    /* Indicate only receive */  
    flag=1;
    break;}
  if(!(relative_rank%mask)){
    '\textit{step}'=mask;
    if(relative_rank+mask>comm_size){
      '\textit{step}'=comm_size-relative_rank;}
    /* Indicate only send */
    flag=0;
    break;}
  mask>>=1;
 }
 
 /* Collect data chunks in (comm_size-1) steps at most */
 for(i=1; i<comm_size; i++){
  rel_j=(j-root+comm_size)%comm_size;
  rel_jnext=(jnext-root+comm_size)%comm_size;
  left_count=$minimum$(scatter_size, (nbytes-rel_jnext*scatter_size));
  if(left_count<0){
    left_count = 0}
  left_disp=rel_jnext*scatter_size;
  right_count=$minimum$(scatter_size, (nbytes-rel_j*scatter_size));
  if(right_count<0){
    right_count=0;}
  right_disp=rel_j*scatter_size;
  
  /* Added code: Jugde if the process has reached the point that indicates either send-only or receive-only */
  if ('\textit{step}'<=comm_size-i){
    MPI_Sendrecv(buffer+right_disp, right_count, MPI_BYTE, right, 0, buffer+left_disp, left_count, MPI_BYTE, left, 0, comm, &status);
  }
  else{
  /* The process reaches the send-only or recevie-only point */
    if(flag){MPI_Recv(buffer+left_disp, left_count, MPI_BYTE, left, 0, comm, &status);}//Receive point
    else{MPI_Send(buffer+right_disp, right_count, MPI_BYTE, right, 0, comm);}//Send point
  }
  
  j=jnext;
  jnext=(comm_size+jnext-1)%comm_size;}
 }
\end{lstlisting}

\section{Experimental Evaluation}
\label{evaluation}
In this section we describe our experiments and give a detailed explanation of the comparison results.
We conducted the micro communication benchmarks for broadcast
to test its bandwidth on two clusters with the following configures for our tests:
\begin{enumerate}
 \item Cray XC40, called Hornet: dual Intel Haswell E5-2680v3 2.5GHz processors compute node with 128GB of main memory, 24 cores per node.
The nodes are interconnected through a Cray Aries network using Dragonfly topology.
 \item NEC Cluster, called Laki: dual Intel Xeon X5560 2.80 GHz processors compute node with 8MB L3 Cache, 8 cores per node.
The nodes are interconnected via Infiniband using switched fabric topology.
\end{enumerate}

All processes are synchronized with a MPI barrier before reaching the broadcast interface.
We then repeat the broadcast operation for 100 iterations and report the bandwidth.
Note, that the bandwidth we present in this section is the rate at which 
the broadcast messages can be processed, and measured in Megabytes per
second (MB/s). Note, that throughout this paper we use megabytes (MB) and
kilobytes (KB) in the base-2 sense, i.e., $2^{20}$ and $2^{10}$, respectively.  

The message size threshold determined by MPICH3 to switch from short messages
to medium messages is 12288 bytes and 
the message size threshold to switch from medium messages to long messages is 524288 bytes.
Thus, we suppose that the long messages should be larger than 
524287 in bytes and medium messages should be larger than 12287 and smaller than 524288 in bytes. 
Our experiments are classified into two types. 
First, we evaluate the tuned design in the case of long message transmission varying the sizes from
524288 to 30000000 bytes with the number of processes of 16, 64 and 256 respectively.
Second, we evaluate the tuned design for medium messages
and long messages with non-power-of-two processes,
as for example 9, 17, 33, 65 and 129 processes.
Third, we measure the tuned design for a range of message sizes from 
12288 to 2560000 bytes with 129 running processes.

Now we introduce two comparison objects for the two experimental platforms -- Hornet and Laki.
For Hornet, we implement the native and tuned broadcast algorithms 
on the user-application level (mentioned in Section \ref{introduce}).
The compiler we used is Cray compiler.
For Laki, we implement the tuned broadcast algorithm on the MPI level, 
which is altered directly inside the MPI code.
%We have integrated the updated broadcast design into the MPICH stack, then configured, compiled and installed it onto the Laki system.
Therefore, the compiler we used is the MPICH in-build compiler -- mpicc.

In this section, we only present the comparison results on Hornet since the results
from both Hornet and Laki basically deliver the same bandwidth performance trend.
\subsection{Long Messages with Power-Of-Two Processes}
\label{pof2}
\begin{figure}
 \begin{center}
  \subfloat[np=16]{\includegraphics[scale=0.65]{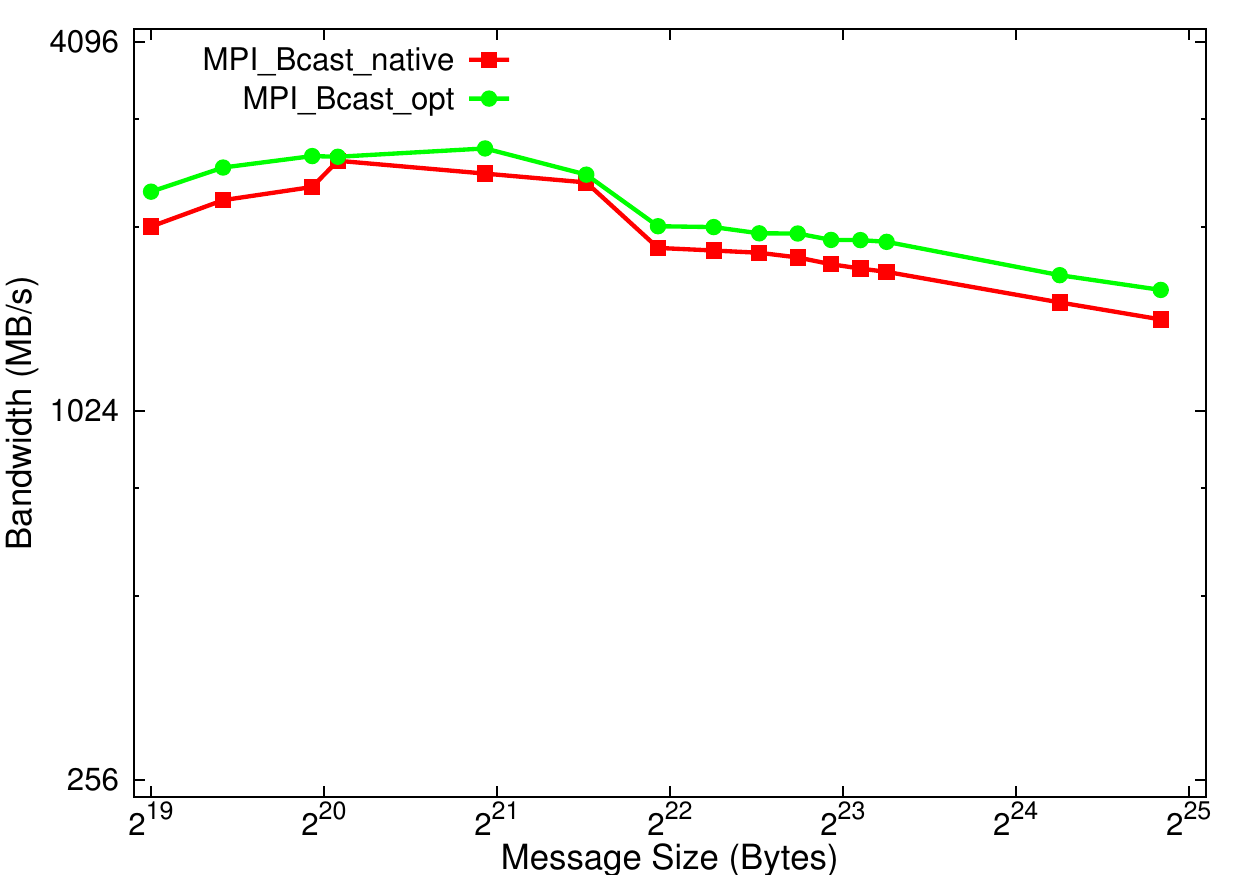} \label{16hornetlmsg}}\\
  \subfloat[np=64]{\includegraphics[scale=0.65]{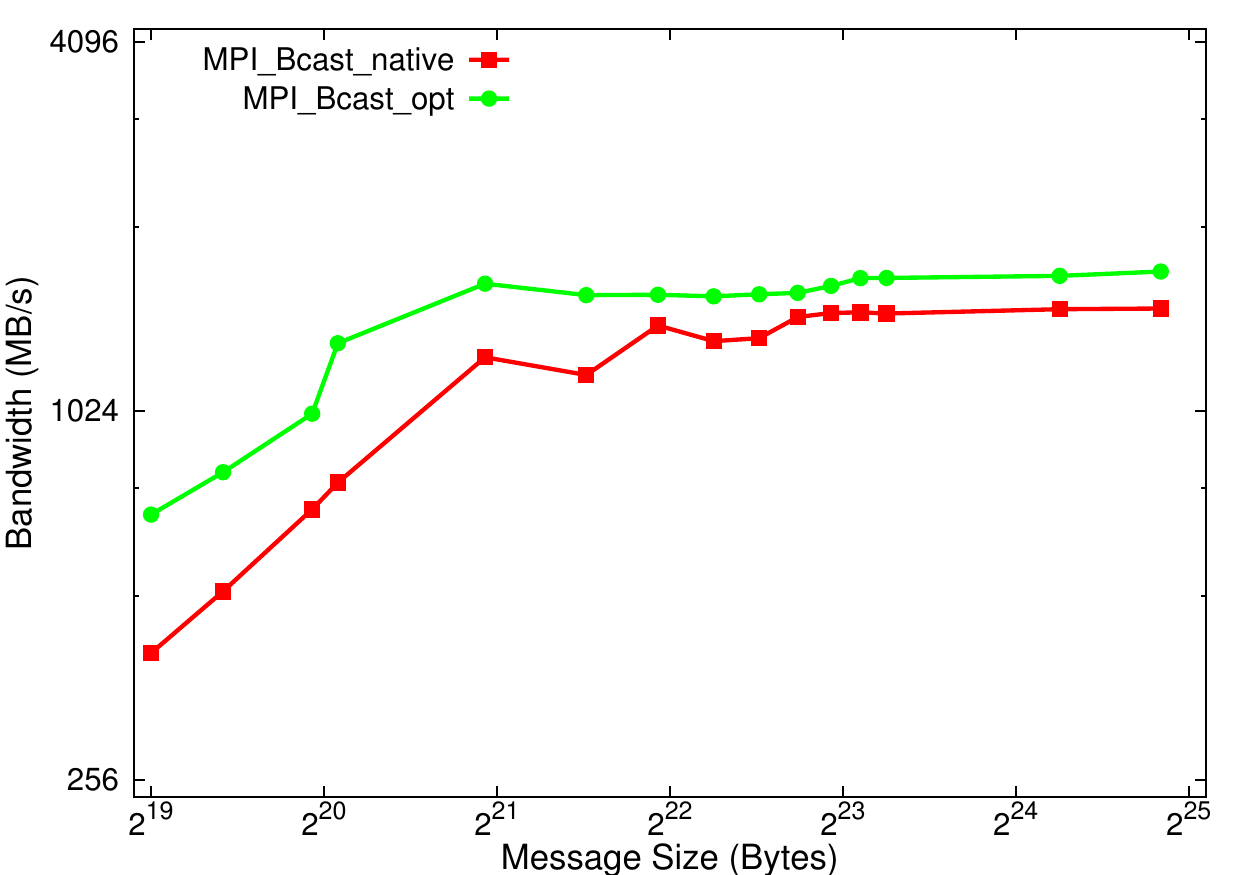} \label{64hornetlmsg}}\\
  \subfloat[np=256]{\includegraphics[scale=0.65]{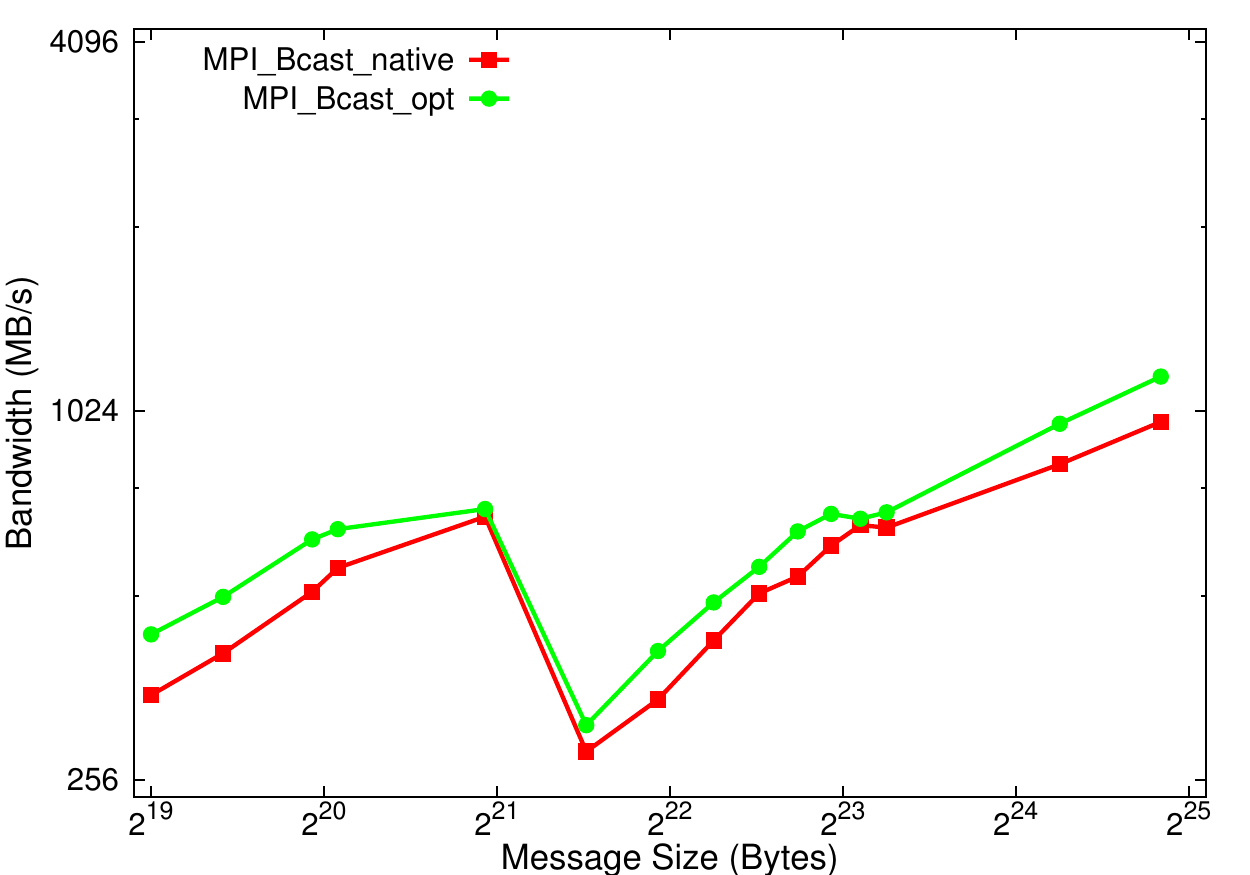} \label{256hornetlmsg}}\\
 \end{center}
 \caption{Bandwidth comparison for long messages with power-of-two processes on Hornet}
 \label{hornetlmsg}
\end{figure}

In this experiment, we measure the bandwidth performance 
of {\em MPI$\_$Bcast\_native} and {\em MPI$\_$Bcast$\_$opt} on Hornet 
over a range of long message sizes with power-of-two processes,
as for example 16, 64 and 256 processes.
All data transmissions occur within one node when only 16 processes are launched since all 
the processes are placed among the nodes in a blocked manner by default on Hornet.
The comparison results are shown in Figure \ref{hornetlmsg} for Hornet system. 
%and \ref{lakilmsg} for Laki system. 
The results are explained as follows:

\indent\textit{16 processes:} As described in Section \ref{design}, 
when the broadcast operation only involve the intra-node data transmission,
the tuned allgather design can avoid extra memory copying operations, which can 
help to save the memory source consumption and alleviate host processing overhead
in the {\em MPI\_Bcast\_opt} implementation. 
The bandwidth performance results for 16 processes are shown
in Figure \ref{hornetlmsg}\protect\subref{16hornetlmsg}.
We observe an improvement of as high as 12\% 
for bandwidth as comparison to {\em MPI$\_$bcast$\_$native} at 512KB, 
The results also show us that the {\em MPI$\_$Bcast$\_$opt} 
consistently outperforms {\em MPI$\_$Bcast$\_$native} even
for very large messages (beyond 10MB).
Additionally,
The {\em MPI\_Bcast\_opt} and
{\em MPI\_Bcast\_native} report a peak bandwidth of up to 2748 MB/s and 2623 MB/s, respectively.
We see that the peak bandwidth performance of {\em MPI\_Bcast\_opt} is 
slightly (10\%) better than that of {\em MPI\_Bcast\_native}.

\indent\textit{64 processes:} When 64 processes are launched on Hornet,
not only is the intra-node data transmission
involved in the broadcast,
but the inter-node data transmission also play a certain role in the broadcast operation.
The growing number of outgoing inter-node messages 
will increase the burden of network routing. 
As can be noted from the Section \ref{design}, compared to \textit{MPI$\_$Bcast$\_$native},
the number of inter-node messages will be reduced to an extent in the {\em MPI$\_$Bcast$\_$opt} algorithm, which leads to
the improvement of the overall broadcast performance in bandwidth.
Further, 
Figure \ref{hornetlmsg}\protect\subref{64hornetlmsg} shows the bandwidth results for 64 processes.
Comparing with the {\em MPI$\_$Bcast$\_$native}
we can observe that the bandwidth achieved by the {\em MPI$\_$Bcast$\_$opt} can be increased by
as high as 41\% for 64 processes on Hornet. 
%However, such advantage degrades slightly as the increase of message size
%as a result of the inherent hardware limitation.
The {\em MPI\_Bcast\_opt} performs 13\% better than the {\em MPI\_Bcast\_native}
in peak bandwidth.

\indent\textit{256 processes:} There are more inter-node messages injecting into network for 256 processes than for 64 processes, Therefore, 
in the case of 256 processes, the performance of the broadcast depends on the network efficiency more than the case of 64 processes
and also the inter-node message transmissions form a greater portion of the overall cost of broadcast operation.
The results shown in Figure \ref{hornetlmsg}\protect\subref{256hornetlmsg} indicate that {\em MPI$\_$Bcast$\_$opt} can yield an
improvement of up to 20\% in bandwidth as comparison to {\em MPI\_Bcast\_native} for 256 processes on Hornet.
Additionally, compared with 
{\em MPI\_Bcast\_native},
the {\em MPI\_Bcast\_opt} improves the peak bandwidth by 16\% for 256 processes.
Moreover, The curves shown in Figure \ref{hornetlmsg}\protect\subref{256hornetlmsg}, Figure \ref{hornetlmsg}\protect\subref{64hornetlmsg} and 
Figure \ref{hornetlmsg}\protect\subref{16hornetlmsg} for 
Hornet suggest that the
{\em MPI\_Bcast\_opt} gains the biggest peak bandwidth advantage for 256 processes.
The drop in bandwidth performance of {\em MPI\_Bcast\_opt} 
and {\em MPI\_Bcast\_ori} starts from around 4MB  for 16 processes, is 
attributed to the limited memory capacity.
Likewise, the bandwidth performance shows slow growth as the increase of transfer message sizes for 64 processes.
We see that there is a sudden drop at around 3MB for 256 processes due to cache effects.

\iffalse
\begin{figure}
 \begin{center}
  \subfloat[np=16]{\includegraphics[scale=0.6]{pic/performance/laki/lakinp16} \label{16lakilmsg}}\\
  \subfloat[np=64]{\includegraphics[scale=0.6]{pic/performance/laki/lakinp64} \label{64lakilmsg}}\\
  \subfloat[np=256]{\includegraphics[scale=0.6]{pic/performance/laki/lakinp256} \label{256lakilmsg}}
 \end{center}
 \caption{Latency Comparison for Long Message with Power-of-Two Processes on Laki}
 \label{lakilmsg}
\end{figure}
\fi
\subsection{Medium Messages and Long Messages with Non-Power-Of-Two Processes}
\iffalse In the implementation of MPICH (shown in the file of \textit{bcast.c}), the medium message is referred to be the message whose size is larger than 12287 in bytes and
meantime smaller than 524288 in bytes, we refer the long message to be the message whose size is larger than 524287 in bytes.
Based on all those data, \fi
In this experiment, we first test 
the throughput (here denoted as the broadcast messages per second) speedups 
of {\em MPI$\_$bcast$\_$opt} over {\em MPI$\_$Bcast$\_$native} for medium messages 
(take two critical message sizes -- 12288 and 524287 bytes for example)
and long messages (take 1048576 bytes for example) with
non-power-of-two processes involved.
Second, we fix the number of processes to 129 and 
then evaluate the bandwidth performance of {\em MPI\_Bcast\_native}
and {\em MPI\_Bcast\_opt} by increasing message sizes from
12288 (medium message size) to 2560000 (long message size) bytes
contiguously.

\begin{figure}
\begin{center}
\includegraphics[scale=0.65]{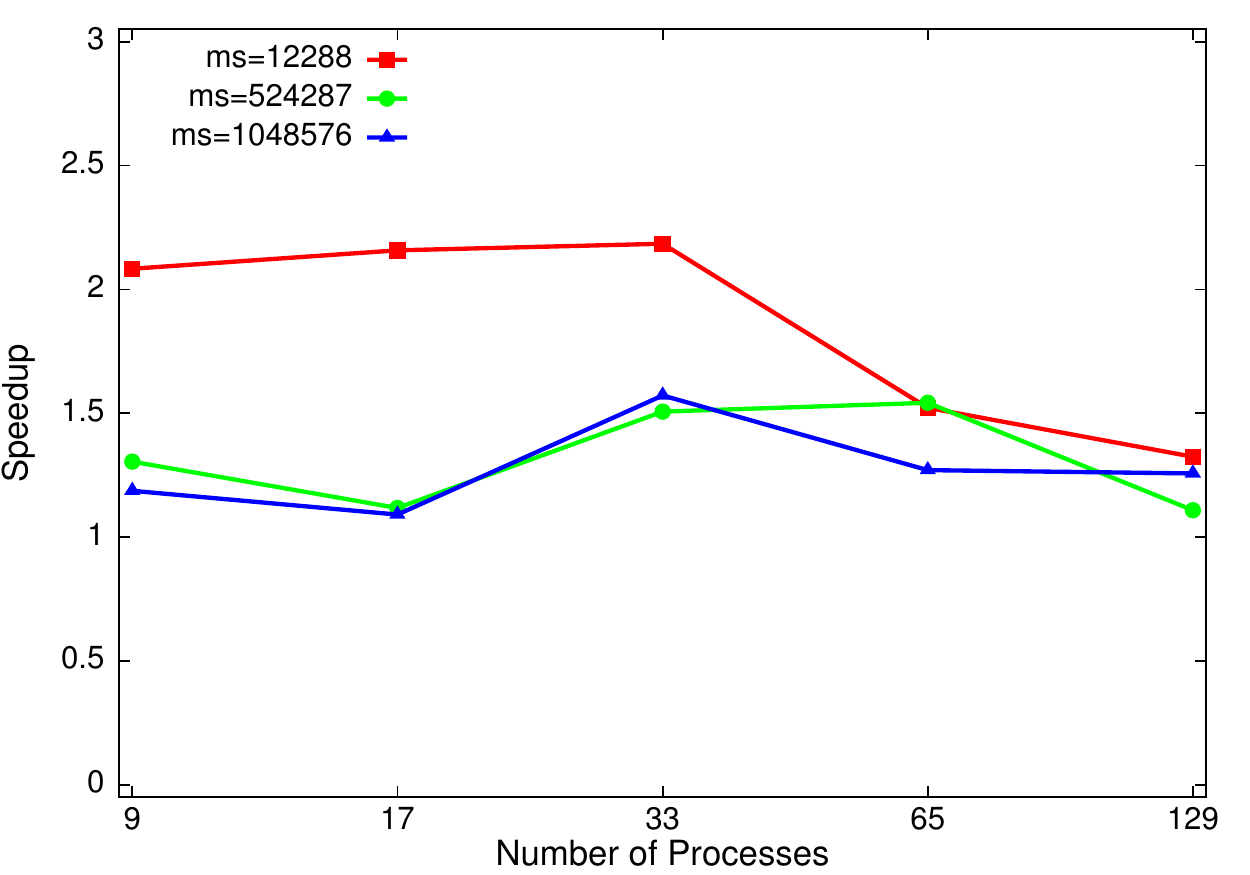}
 \end{center}
 \caption{Throughput speedups of {\em MPI$\_$Bcast$\_$opt} over {\em MPI$\_$Bcast$\_$native}}
 \label{speedup}
\end{figure}

Figure \ref{speedup} shows the throughput speedups of {\em MPI$\_$Bcast$\_$opt} over {\em MPI$\_$Bcast$\_$native}. 
Specifically, the significantly higher speedups are achieved for message size of 12288 bytes
than for the other two cases -- message sizes of 524287 bytes and 1048576 bytes.
We can see that {\em MPI$\_$Bcast$\_$opt} performs more than two times better than {\em MPI\_Bcast\_native} for 12288 bytes
in the case of 9, 17 and 33 processes.
However, as can be noted from the speedup trend in Figure \ref{speedup}, the speedup goes down suddenly as the transferred message sizes are 
increased on up to 65 processes for 12288 bytes.
Regarding the case of 524287 bytes and 1048576 bytes, they show similar speedups on the measured number of processes -- 9, 17, 33 and 129.
In addition, compared with the case of 12288 bytes
and 524287 bytes, a more stable speedup curve is presented for 1048576 bytes.
In overall, we can observe that in the case of non-power-of-two processes the {\em MPI\_Bcast\_opt} consistently performs better 
than {\em MPI\_Bcast\_native} in throughput no matter what ratio of 
inter-node messages to intra-node messages involved in the broadcast operation is.

\begin{figure}
\begin{center}
\includegraphics[scale=0.65]{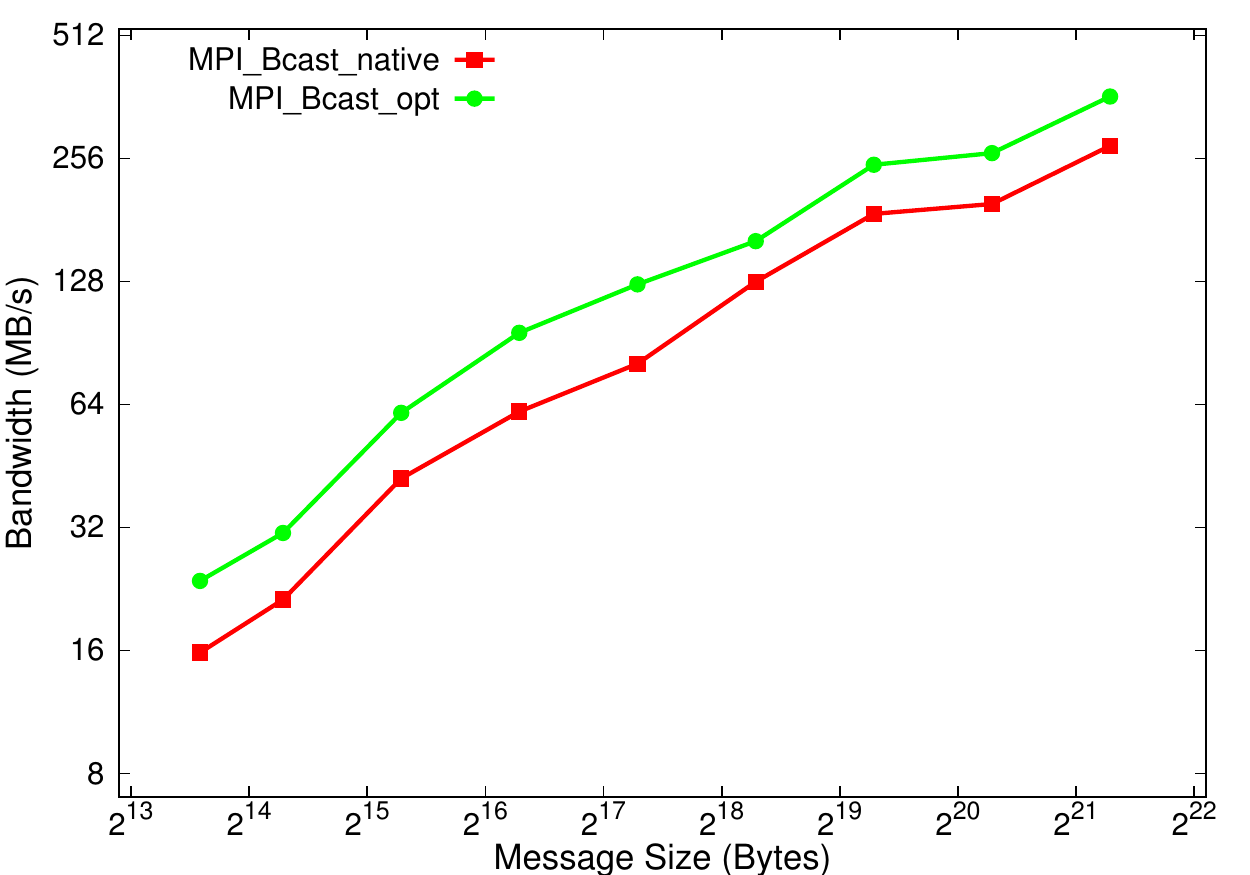}
 \end{center}
 \caption{Bandwidth comparison for medium and long messages with 129 processes on Hornet}
 \label{hornetnpow2}
\end{figure}

Figure \ref{hornetnpow2} takes, 129 processes for example,
and shows the variance in bandwidth as the message sizes are increased
from 12288 to 2560000 bytes.
From this figure, we can conclude that the bandwidth increases steadily
as the growth of message sizes under conditions that have sufficient
memory capacity.
In the best case, the bandwidth achieved by {\em MPI\_Bcast\_opt}
get improved by up to 30\% as comparison to {\em MPI\_Bcast\_ori}.
No sudden change is expected in the curves shown in Fig.~\ref{hornetnpow2}
since Cray MPI keeps using rendezvous message protocol
for message sizes ranging from 12288 to 2560000 bytes.

\section{Conclusion}
\label{conclusion}
In this paper, we have proposed an optimized design for the MPI broadcast collective operation based on the native broadcast
algorithm used by MPICH3. We observe that, in the case of long messages or medium messages but non-power-of-two process counts,
the suboptimal ring allgather algorithm used by broadcast operation
in MPICH3 leads to the inefficiency of broadcast operation.
Therefore, we designed a tuned ring allgather algorithm to serve MPI broadcast implementation.
Our design aims to reduce the amount of data transmission traffic brought by the native ring allgather operation, which in turn 
eases the burden of network and host processing.
Additionally, our performance evaluation at the user level on current Cray system reveals that
the tuned broadcast algorithm can reduce the bandwidth by a range from 2\% to 41\% for long messages with power-of-two processes
(16, 64 and 256 processes).
For non-power-of-two processes, 
the tuned design is also superior to the native one 
according to the throughput speedup curve.
Additionally, with 129 running processes,
the bandwidth achieved by the tuned broadcast design
get improved by 30\% at best. 

%, and the associated speedups look relatively scalable.  


\section*{Acknowledgment}
This work has been supported by the CRESTA project funded by
 the European Community's Seventh Framework Programme
(ICT-2011.9.13) under Grant Agreement no.287703. We gratefully
acknowledge funding by the German Research
Foundation (DFG) through the German Priority Programme 1648 Software
for Exascale Computing (SPPEXA).

%more thanks here



%\bibliographystyle{IEEEtran}
\bibliographystyle{IEEEtran}
\bibliography{IEEEabrv,longmegopt}

\begin{thebibliography}{10}
\providecommand{\url}[1]{#1}
\csname url@samestyle\endcsname
\providecommand{\newblock}{\relax}
\providecommand{\bibinfo}[2]{#2}
\providecommand{\BIBentrySTDinterwordspacing}{\spaceskip=0pt\relax}
\providecommand{\BIBentryALTinterwordstretchfactor}{4}
\providecommand{\BIBentryALTinterwordspacing}{\spaceskip=\fontdimen2\font plus
\BIBentryALTinterwordstretchfactor\fontdimen3\font minus
  \fontdimen4\font\relax}
\providecommand{\BIBforeignlanguage}[2]{{%
\expandafter\ifx\csname l@#1\endcsname\relax
\typeout{** WARNING: IEEEtran.bst: No hyphenation pattern has been}%
\typeout{** loaded for the language `#1'. Using the pattern for}%
\typeout{** the default language instead.}%
\else
\language=\csname l@#1\endcsname
\fi
#2}}
\providecommand{\BIBdecl}{\relax}
\BIBdecl

\bibitem{mpi}
``{The Message Passing Interface (MPI) standard},'' [online],
  http://www.mcs.anl.gov/research/projects/mpi/.

\bibitem{mpi3}
{MPI Forum}, ``{\textsf{MPI}: A Message-Passing Interface Standard. Version
  3.0},'' September 21st 2012, available at:
  \url{http://www.mpi-forum.org/docs/mpi-3.0/mpi30-report.pdf} (Sept. 2012).

\bibitem{hpl}
A.~Petitet, R.~Whaley, J.~Dongarra, and A.~Cleary, ``{HPL - A Portable
  Implementation of the High-Performance Linpack Benchmark for
  Distributed-Memory Computers},'' [online],
  http://www.netlib.org/benchmark/hpl/.

\bibitem{LS-DYNA}
G.~Shainer, T.~Liu, P.~Lui, and D.~Field, ``{The Effect of MPI Collective
  Operations and MPI Collective Offloads on LS-DYNA Performance},'' in
  \emph{{8th European LS-DYNA ® Users Conference}}, May 2011.

\bibitem{mpich-overview}
``{MPICH Overview},'' [online], http://www.mpich.org/about/overview/.

\bibitem{top500}
``Top500 - list statistics - november 2014,''
  \url{http://www.top500.org/statistics/list/}, accessed: June 2015.

\bibitem{multicore}
L.~Chai, Q.~Gao, and D.~K. Panda, ``{Understanding the Impact of Multi-Core
  Architecture in Cluster Computing: A Case Study with Intel Dual-Core
  System.}'' in \emph{{CCGRID}}.\hskip 1em plus 0.5em minus 0.4em\relax IEEE
  Computer Society, 2007, pp. 471--478.

\bibitem{bcast-tu}
\BIBentryALTinterwordspacing
B.~Tu, M.~Zou, J.~Zhan, X.~Zhao, and J.~F. 0002, ``{Multi-core aware
  optimization for MPI collectives.}'' in \emph{{CLUSTER}}.\hskip 1em plus
  0.5em minus 0.4em\relax IEEE, 2008, pp. 322--325. [Online]. Available:
  \url{http://ieeexplore.ieee.org/xpl/mostRecentIssue.jsp?punumber=4655410}
\BIBentrySTDinterwordspacing

\bibitem{mpich-opt}
R.~Thakur, R.~Rabenseifner, and W.~Gropp, ``{Optimization of Collective
  Communication Operations in MPICH.}'' \emph{IJHPCA}, vol.~19, no.~1, pp.
  49--66, 2005.

\bibitem{mpich-sc}
``{MPICH Source Code},'' [online], http://www.mpich.org/downloads/.

\bibitem{bcast-multicast-liu}
J.~Liu, A.~Mamidala, and D.~Panda, ``{Fast and scalable MPI-level broadcast
  using InfiniBand's hardware multicast support},'' in \emph{{Parallel and
  Distributed Processing Symposium, 2004. Proceedings. 18th International}},
  April 2004, pp. 10--.

\bibitem{bcast-hoefler}
T.~Hoefler, C.~Siebert, and W.~Rehm, ``{A practically constant-time MPI
  Broadcast Algorithm for large-scale InfiniBand Clusters with Multicast.}'' in
  \emph{{IPDPS}}.\hskip 1em plus 0.5em minus 0.4em\relax IEEE, 2007, pp. 1--8.

\bibitem{conf/europar/bcast_sdn}
K.~Dashdavaa, S.~Date, H.~Yamanaka, E.~Kawai, Y.~Watashiba, K.~Ichikawa,
  H.~Abe, and S.~Shimojo, ``{Architecture of a High-Speed MPI\_Bcast Leveraging
  Software-Defined Network.}'' in \emph{{Euro-Par Workshops}}, ser. {Lecture
  Notes in Computer Science}, vol. 8374.\hskip 1em plus 0.5em minus 0.4em\relax
  Springer, 2013, pp. 885--894.

\bibitem{journals/ijhpca/bcast_ibm}
S.~Kumar, A.~R. Mamidala, P.~Heidelberger, D.~Chen, and D.~Faraj,
  ``{Optimization of MPI collective operations on the IBM Blue Gene/Q
  supercomputer.}'' \emph{IJHPCA}, vol.~28, no.~4, pp. 450--464, 2014.

\end{thebibliography}

\end{document}